# Models of Flow through Porous Media of Polar Fluids with Pressure-Dependent Viscosity

**M.H. HAMDAN**[*], **FIMA**
Department of Mathematics and Statistics
hamdan@unb.ca
*Corresponding Author

**D.C. ROACH**
Department of Engineering
droach@unb.ca

University of New Brunswick
100 Tucker Park Road, Saint John, New Brunswick
CANADA E2L 4L5

*Abstract:* - Equations governing the flow of a polar fluid, with pressure-dependent Newtonian viscosity, through a variable-porosity medium are developed. Averaged equations are obtained using intrinsic volume averaging. A drag function is introduced to account for the interactions of the fluid with the porous matrix. Darcy and Forchheimer generalized terms are included in the model equations for both granular and consolidated media to account for the effects of the porous microstructure.

## 1 Introduction

Several theories have been developed to describe the motion of classes of fluids possessing microstructures. Cowin's polar fluid theory [1] is amongst the simplest and has been approached from different perspectives. It is recognized under various acronyms that include asymmetric hydrodynamics [2] and theory of micro-polar fluids [3]. Flow of polar fluids has received considerable attention in the scientific literature due its spectrum of important applications in physical and biological systems that include applications in fields such as lubrication theory, blood flow, flow of oil and gas, heat and mass transfer, energy studies and industrial designs, polymer transport, microfluidics, and in the study of nanofluids (*cf.* [4], [5], [6], [7], [8], [9], [10], [11], [12], [13] and the references therein.)

Motion of a polar fluid is described by two independent kinematic vector fields: a Newtonian velocity vector field and an axial vector that models spin or micro-rotation of the (assumingly rigid) fluid particles. Angular effects, such as couple stress and asymmetric stress tensor, are present, and the governing equations consist of an equation of continuity, a linear momentum and an angular momentum equation. Solutions to the governing equations in the presence of solid boundaries require imposing appropriate boundary conditions. The no-slip condition for the velocity and the no-spin condition for the micro-rotation vector have been widely used, [1], [8], [14], [15], [16]. Governing equations take the following form when the flow is steady, the fluid is incompressible, body forces and body couple are absent, [1], [3]:

$$\nabla \cdot \vec{v} = 0 \qquad (1)$$

$$\rho \nabla \cdot \vec{v}\vec{v} = -\nabla p + (\eta + \tau)\nabla^2 \vec{v} + 2\tau \nabla \times \vec{G} \qquad (2)$$

$$\rho k^2 \nabla \cdot \vec{v}\vec{G} = 2\tau \nabla \times \vec{v} + (\alpha + \beta - \gamma)\nabla(\nabla \cdot \vec{G}) + (\beta + \gamma)\nabla^2 \vec{G} - 4\tau\vec{G} \tag{3}$$

where $\vec{v}$ is the velocity vector field, $\vec{G}$ is the microrotation vector, $p$ is the pressure, $\rho$ is the fluid density, $k$ is the radius of gyration of a fluid element, in a given volume element, about the centroid of the volume element, $\eta$ is the fluid viscosity (Newtonian), $\tau$ is the rotational viscosity, and $\alpha, \beta, \gamma$ are gradient viscosities.

In modelling flow of a polar fluid in a porous medium, complexity of the pore geometry and the absence of a mathematical description of the solid matrix, together with the existence of various types of porous media, renders description of fluid flow (including polar fluid flow) at the microscopic, pore level a formidable task. This has given rise to homogenization methods and averaging techniques as means of describing polar fluid flow through porous media. Homogenization methods are popular in the modelling of polymer flow. [17] presented analyses and homogenization of a polymer flow through a porous microstructure and emphasized the nonlinear nature of the viscosity of the fluid as a function of the rate of strain tensor. Homogenization of a micropolar fluid flow through a porous structure was presented by [18] who also investigated the limit flow of a non-stationary micropolar fluid in a periodic porous structure.

Volume averaging techniques have gained popularity in describing the flow through porous structures since the introduction of the averaging theorem, [19]. In one aspect, intrinsic volume averaging allows for flow modelling through porous media of variable characteristics (such as permeability and porosity) and allows for considering the porous microstructure in modelling solid matrix effects and resistance to the flow. In applying intrinsic volume averaging, [20], [21], [22] modelled the porous microstructure effects as a Darcy resistance and a Forchheimer quadratic inertial term, in the linear momentum equations. Effects of the couple stress on the flow and heat transfer in porous media was first analysed by [23], [24], [25], [26], by employing a generalized Darcy's resistance in their analysis (generalized in the sense that the Darcy resistance incorporates both the fluid viscosity and the rotational viscosity). Subsequently, [27] considered the generalized Forchheimer term, while [28] incorporated both the generalized Darcy and Forchheimer terms in his study of boundary layer flow of a micropolar fluid through a porous medium.

It is worth noting that considering the porous microstructure in a flow model renders the Darcy resistance term a deciding factor in the form the governing equations take for a given porous microstructure. In other words, a distinguishing feature between two models governing the laminar flow of a polar fluid through two different porous media is the form the Darcy resistance takes. [20] obtained the following equations by intrinsically averaging equations (1), (2) and (3), under the assumption of a non-zero divergence of the microrotation vector, namely $\nabla \cdot \vec{G} \neq 0$:

*Continuity Equation:*

$$\nabla \cdot \varphi <\vec{v}>_\varphi = 0 \tag{4}$$

*Linear Momentum Equation:*

$$\rho \nabla \cdot \varphi <\vec{v}>_\varphi <\vec{v}>_\varphi = -\varphi \nabla <p>_\varphi + (\eta + \tau)\nabla^2 \varphi <\vec{v}>_\varphi + 2\tau \nabla \times \varphi <\vec{G}>_\varphi - \mu f <\vec{v}>_\varphi \tag{5}$$

where $f$ is the Churchill-Usagi total frictional effects of the porous matrix on the fluid. It is customary, however, to write $\frac{f}{\varphi}<\vec{v}>_\varphi = -\frac{(\eta+\tau)}{\sigma}<\vec{v}>_\varphi$ to represent Darcy resistance to the flow, wherein $\sigma$ is permeability of the medium.

*Angular Momentum Equation:*

$$\rho k^2 \nabla \cdot \varphi <\vec{v}>_\varphi <\vec{G}>_\varphi = 2\tau \nabla \times \varphi <\vec{v}>_\varphi + (\alpha + \beta - \gamma)\varphi \nabla \left(\frac{1}{\varphi}\nabla \cdot \varphi <\vec{G}>_\varphi\right) + (\beta + \gamma)\nabla^2 \varphi <\vec{G}>_\varphi - 4\tau\varphi <\vec{G}>_\varphi \tag{6}$$

Of interest in many applications is the flow of a polar fluid under the effects of high pressures. This type of flow necessitates developing model equations that consider viscosity variations due to pressure. These variations, in the case of Newtonian fluid flow, have been reported it the experiments of Barus, [29], who concluded that fluid viscosity is an exponential function of pressure of the form $\eta(p) = \eta_0 e^{\delta p}$, where $\eta_0$ is a reference constant viscosity and $\delta$ is a control parameter. A question that arises in connection with viscosity variations is: which viscosity is affected by pressure? The simple answer to this question is that rotational viscosity is not dependent on pressure! This is explained in what follows.

In their investigation of the rotational viscosity for a chlorine fluid and for a fluid composed of small linear molecules by using equilibrium molecular dynamics simulations, [30] found that the rotational viscosity is almost independent of temperature but exhibits a power-law dependency on density. In their evaluation of the rotational viscosity and the two spin viscosities for liquid water using equilibrium molecular dynamics, [31] found that the rotational viscosity is independent of the temperature in the range from $284° - 319°$ K, while the two spin viscosities decrease with increasing temperature. [32] studied the effects of viscosity variations of micropolar flow of nanofluids. They considered variations in dynamic viscosity, spin-gradient viscosity, and micro-inertia density, to inversely vary with temperature. [6], assumed that spin viscosity, Newtonian viscosity and viscosity coefficient for micropolar fluids all vary with pressure in the same way.

The scope of the current work is to develop model equations of flow through porous media that consider the dependence of viscosity on pressure. At the outset, it is assumed that Newtonian viscosity varies with pressure, while other viscosities are taken as constant. To accomplish this, the method of intrinsic volume averaging will be used, and the effects of the porous microstructure will be accounted for using the concept of a representative unit cell. Both Darcy and Forchheimer effects are described using generalized Darcy and generalized Forchheimer models, proposed by [33] and [34]. This work is based on the work reported in [35] and represents an expanded form of it.

## 2 Models Development

Letting $\mu = \eta + \tau$, and assuming that $\eta$ is variable and $\tau$ is constant, and that $\mu = \tau + \eta(p)$, equation (2) can be written in the form:

$$\rho \nabla \cdot \vec{v}\vec{v} = -\nabla p + \nabla \cdot \vec{T} + 2\tau \nabla \times \vec{G} \qquad (7)$$

wherein

$$\vec{T} = \mu(\nabla \vec{v} + (\nabla \vec{v})^T) \qquad (8)$$

Derivation of the equations governing the flow of a polar fluid with pressure-dependent viscosity through a porous medium involves intrinsically averaging equations (1), (3) and (7) over a representative elementary volume, REV. The averaging procedure and averaging rules are given in **Appendix 1.** The averaged forms of (1) and (3) are the same as given by (4) and (6). In what follows, only the linear momentum equation (7) will be averaged according to rules of averaging in **Appendix 1.** Typical conditions on the velocity and spin vectors are the no-slip and no-spin assumptions on the solid matrix. These are implemented in this work and translate into: $\vec{v} = \vec{G} = \vec{0}$ on the stationary solid matrix.

Term-by-term averaging of equation (7) gives:

$$\rho \nabla \cdot \varphi <\vec{v}>_\varphi <\vec{v}>_\varphi +$$

$$\rho \nabla \cdot \varphi <\vec{v}°\vec{v}°>_\varphi + \frac{\rho}{V}\int_S \vec{v}\vec{v}\cdot\vec{n}dS =$$

$$-\varphi \nabla <p>_\varphi - \frac{1}{V}\int_S p°\vec{n}dS \qquad (9)$$

$$+\nabla \cdot \varphi <\vec{T}>_\varphi + \frac{1}{V}\int_S \vec{T}\vec{n}dS +$$

$$2\tau \nabla \times \varphi <\vec{G}>_\varphi + \frac{2\tau}{V}\int_S \vec{G}\times\vec{n}\,dS$$

### 2.1 Analysis of the Deviation Terms and Surface Integrals

The surface integrals and deviation terms in equation (9) embed the information needed to analyse the interactions of the fluid and the porous matrix. The no-slip and no-spin conditions on solid matrix walls imply that the terms involving velocity and microrotation vectors explicitly, vanish. Accordingly, the surface integrals $\int_S v\vec{v}\cdot\vec{n}dS$ and $\int_S \vec{G}\times\vec{n}dS$ vanish.

The volume filter $\nabla \cdot \varphi <\vec{v}°\vec{v}°>_\varphi$ contains the hydrodynamic dispersion of the average velocity. Hydrodynamic dispersion through porous media is the sum of mechanical dispersion due to tortuosity of the flow path within the porous microstructure, and molecular diffusion arising from diffusion of fluid vorticity. [36], [37], analysed the velocity deviation term and concluded that it is negligible in the absence of high porosity and velocity gradients. [38] justified ignoring terms involving the divergence of the product of porosity and velocity deviations on the bases of uniform average flow through porous media of uniform porosities. Accordingly, the above volume filters are ignored in equation (9). Equation (9) thus reduces to the following form:

$$\rho \nabla \cdot \varphi <\vec{v}>_\varphi <\vec{v}>_\varphi = -\varphi \nabla <p>_\varphi \\ - \frac{1}{V}\int_S p°\vec{n}dS + \nabla \cdot \varphi <\vec{T}>_\varphi \\ + \frac{1}{V}\int_S \vec{T} \cdot \vec{n}dS + 2\tau \nabla \times \varphi <\vec{G}>_\varphi \quad (10)$$

The term $\frac{1}{V}\int_S p°\vec{n}dS$ represents pressure fluctuations on the fluid-solid interface. In case of flow of a Newtonian fluid with constant viscosity, it can be argued that this term is small, hence can be neglected, (*cf.* [39] and the references therein). However, in case of high-pressure flow through porous media, it may be of significance.

Surface integrals $\frac{1}{V}\int_S p°\vec{n}dS$ and $\frac{1}{V}\int_S \vec{T}\cdot\vec{n}dS$ of equation (10), can be combined to form a *surface filter*, $\frac{1}{V}\int_S (\overline{\vec{T}\cdot\vec{n}} - p°\vec{n})dS$, that involves the normal derivative of $\vec{v}$. Since the solid porous matrix affects the fluid through the portion of the surface area of the solid that is in contact with it, this surface filter contains the information necessary to quantify the forces exerted on the flowing fluid by the porous matrix.

The term $\frac{1}{V}\int_S \vec{T}\cdot\vec{n}dS$ is the interfacial viscous stress exchange, which corresponds to the microscopic momentum exchange of the Newtonian fluid with the solid matrix. It depends on the morphology of the porous medium and, if present, on the relative velocity of the fluid and the solid matrix.

## 2.2 Analysis of Microstructural Effects

The above surface filter has been analysed in the literature and has been identified with the force that gives rise to Darcy resistance and the Forchheimer inertial drag term, [39]. The surface filter can be expressed in the following form:

$$\frac{1}{V}\int_S -(p°\vec{n} - \vec{T}\cdot\vec{n})dS \\ = -\varphi <\mu>_\varphi f <\vec{v}>_\varphi \quad (11)$$

wherein the average viscosity is a function of pressure, and $f$ is a friction factor (or the Churchill-Usagi total frictional effects of the porous matrix on the fluid, [38]). To account for Darcy resistance and the Forchheimer drag effects, the surface integral is decomposed as follows (*cf.* [22], and the references therein):

1) A shear force integral, $f_1$, which accounts for the viscous drag effects (Darcy resistance) that predominate in the Darcy regime, that is, for small Reynolds number flow.
2) An inertial force integral, $f_2$, which accounts for inertial drag effects that predominate in the Forchheimer regime, that is, for high Reynolds number flow.
3) If both $f_1$ and $f_2$ are accounted for, then, by the Churchill-Usagi relation, total frictional effects, $f$, of the porous matrix on the fluid is expressed in terms of $f_1$ and $f_2$ as

$$f^r = f_1^r + f_2^r \quad (12)$$

where $r$ is a shifting factor. [38] showed that reasonable correlation is obtained when $r = 1$. We thus write $f = f_1 + f_2$, where $f_1$ be the *velocity-independent* viscous shear geometric factor that depends on the geometry of the porous medium and gives rise to the Darcy resistance, and $f_2$ the *velocity-dependent* inertial geometric factor that gives rise to the Forchheimer inertial term.

Expressions for $f_1$ and $f_2$ require a mathematical description of the porous matrix and its microstructure. [38] carried out extensive analysis on evaluating these geometric factors for isotropic porous media, based on the concept of a Representative Unit Cell (RUC), which they defined as the minimal REV in which the average properties of the porous medium are embedded, [36], [37]. For granular and consolidated isotropic porous media, the following expressions, summarized in **Table 1**, as given in [38], are adopted in this work for $f_1$ and $f_2$ and for the hydrodynamic permeability, $\sigma$:

**Table 1. Granular and Consolidated Media Friction Factors**

| Granular Media | Consolidated Media |
|---|---|
|  |  |
| $f_1 = \dfrac{36(1-\varphi)^{\frac{2}{3}}}{d^2[1-(1-\varphi)^{\frac{1}{3}}][1-(1-\varphi)^{\frac{2}{3}}]}$ | $f_1 = \dfrac{36\epsilon(\epsilon-1)}{d^2\varphi}$ |
| $f_2 = \dfrac{\rho d\|\varphi<\vec{v}>_\varphi\|C_d(1-\varphi)^{\frac{2}{3}}}{d^2<\mu>_\varphi[1-(1-\varphi)^{\frac{2}{3}}]^2}$ | $f_2 = \dfrac{\rho d\|\varphi<\vec{v}>_\varphi\|C_d\epsilon(\epsilon-1)}{d^2<\mu>_\varphi\varphi(3-\epsilon)}$ |
| $\sigma = \dfrac{\varphi}{f_1}$ $= \dfrac{d^2\varphi[1-(1-\varphi)^{\frac{1}{3}}][1-(1-\varphi)^{\frac{2}{3}}]}{36(1-\varphi)^{\frac{2}{3}}}$ | $\sigma = \dfrac{\varphi}{f_1} = \dfrac{d^2\varphi^2}{36\epsilon(\epsilon-1)}$ |

where $d$ is a microscopic length (such as the mean pore diameter) and $C_d$ is the Forchheimer drag coefficient, and tortuosity, $\epsilon$, is approximated by $\epsilon \approx \dfrac{1+2\varphi}{3}$.

The linear momentum equation (10) can be written in the following forms after substituting the expressions for the friction factors from **Table 1**.

*For Granular Media:*

$$\rho\nabla\cdot\varphi<\vec{v}>_\varphi<\vec{v}>_\varphi = -\varphi\nabla<p>_\varphi$$
$$+\nabla\cdot\varphi<\vec{T}>_\varphi + 2\tau\nabla\times\varphi<\vec{G}>_\varphi$$
$$-\dfrac{36(1-\varphi)^{\frac{2}{3}}\varphi<\mu>_\varphi}{d^2[1-(1-\varphi)^{\frac{1}{3}}][1-(1-\varphi)^{\frac{2}{3}}]}<\vec{v}>_\varphi \qquad (13)$$
$$-\dfrac{\rho C_d(1-\varphi)^{\frac{2}{3}}\varphi}{d[1-(1-\varphi)^{\frac{2}{3}}]^2}<\vec{v}>_\varphi\,|\varphi<\vec{v}>_\varphi|$$

*For Consolidated Media:*

$$\rho\nabla\cdot\varphi<\vec{v}>_\varphi<\vec{v}>_\varphi = -\varphi\nabla<p>_\varphi$$
$$+\nabla\cdot\varphi<\vec{T}>_\varphi + 2\tau\nabla\times\varphi<\vec{G}>_\varphi \qquad (14)$$
$$-\dfrac{36\epsilon(\epsilon-1)}{d^2}<\mu>_\varphi<\vec{v}>_\varphi$$
$$-\dfrac{\rho C_d\epsilon(\epsilon-1)}{d(3-\tau)}<\vec{v}>_\varphi\,|\varphi<\vec{v}>_\varphi|$$

Now, letting

$$\vec{q} = \varphi<\vec{v}>_\varphi,$$
$$\vec{g} = \varphi<\vec{G}>_\varphi,$$
$$p^* = <p>_\varphi,$$
$$\mu^* = \varphi<\mu>_\varphi = \varphi<\eta+\tau>_\varphi$$
$$= \varphi<\eta>_\varphi + \tau = \eta_0 e^{\delta p^*} + \tau,$$
$$\vec{T}^* = \varphi<\vec{T}>_\varphi = \mu^*\left\{\nabla\left(\dfrac{\vec{q}}{\varphi}\right) + \nabla\left(\dfrac{\vec{q}}{\varphi}\right)^T\right\},$$

equations (13) and (14) take the following forms, respectively:

$$\rho\nabla\cdot\vec{q}\vec{q}/\varphi = -\varphi\nabla p^* + \nabla\cdot\vec{T}^* + 2\tau\nabla\times\vec{g}$$
$$-\dfrac{36(1-\varphi)^{\frac{2}{3}}(\eta_0 e^{\delta p^*}+\tau)}{d^2\varphi[1-(1-\varphi)^{\frac{1}{3}}][1-(1-\varphi)^{\frac{2}{3}}]}\vec{q} \qquad (15)$$
$$-\dfrac{\rho C_d(1-\varphi)^{\frac{2}{3}}}{d[1-(1-\varphi)^{\frac{2}{3}}]^2}\vec{q}|\vec{q}|$$

$$\rho\nabla\cdot\dfrac{\vec{q}\vec{q}}{\varphi} = -\varphi\nabla p^* + \nabla\cdot\vec{T}^* + 2\tau\nabla\times\vec{g}$$
$$-\dfrac{36\epsilon(\epsilon-1)}{(\varphi d)^2}(\eta_0 e^{\delta p^*}+\tau)\vec{q} \qquad (16)$$
$$-\dfrac{\rho C_d\epsilon(\epsilon-1)}{d\varphi(3-\epsilon)}\vec{q}|\vec{q}|$$

Equations (15) and (16) can alternatively be obtained by expressing the surface filter in the following form:

$$\frac{1}{V}\int_S -(p°\vec{n} - \vec{T}\cdot\vec{n})dS = -\alpha(p^*)\varphi <\vec{v}>_\varphi \quad (17)$$

and writing the linear momentum as

$$\rho\nabla\cdot\frac{\vec{q}\vec{q}}{\varphi} = -\varphi\nabla p^* + \nabla\cdot\vec{T}^* \\ +2\tau\nabla\times\vec{g} - \alpha(p^*)\vec{q} \quad (18)$$

where $\alpha(p^*)$ is a pressure-dependent drag coefficient. In [33], it is argued that the fluid viscosity is a measure of frictional resistance in between fluid layers while $\alpha(p^*)$ is a measure of the friction between the fluid and the solid, at the pore. [34] provided generalized versions of Darcy's and Forchheimer's equations using drag coefficients. Using the expressions of **Table 1**, the following forms of drag coefficients are obtained:

*For Granular Media*

*Darcy's Drag Coefficient:*

$$\alpha(p,\vec{x}) = \frac{36(1-\varphi)^{\frac{2}{3}}(\eta_0 e^{\beta_B p^*} + \tau)}{d^2\varphi[1-(1-\varphi)^{\frac{1}{3}}][1-(1-\varphi)^{\frac{2}{3}}]} \quad (19)$$

*Forchheimer Drag Coefficient:*

$$\alpha(\vec{v},p,\vec{x}) = \frac{36(1-\varphi)^{\frac{2}{3}}(\eta_0 e^{\beta_B p^*} + \tau)}{d^2[1-(1-\varphi)^{\frac{1}{3}}][1-(1-\varphi)^{\frac{2}{3}}]} \\ + \frac{\rho C_d(1-\varphi)^{\frac{2}{3}}}{d[1-(1-\varphi)^{\frac{2}{3}}]^2}|\vec{q}| \quad (20)$$

*For Consolidated Media*

*Darcy's Drag Coefficient:*

$$\alpha(p,\vec{x}) = \frac{36\tau(\tau-1)}{(\varphi d)^2}(\eta_0 e^{\beta_B p^*} + \tau) \quad (21)$$

*Forchheimer Drag Coefficient:*

$$\alpha(\vec{v},p,\vec{x}) = \frac{36\tau(\tau-1)}{(\varphi d)^2}\cdot(\eta_0 e^{\beta_B p^*} + \tau) \\ + \frac{\rho C_d\tau(\tau-1)}{d\varphi(3-\tau)}|\vec{q}| \quad (22)$$

If $\beta_F$ is identified with $\delta$, then using (20) and (22) in equation (18) yields (15) and (16), respectively.

## 3 Final Forms of Governing Equations

Equations governing the flow of a polar fluid, with pressure-dependent fluid viscosity, through a variable porosity medium are given by equations (4), (6), and (15) or (16). These are written in the following final forms:

*Continuity Equation:*

$$\nabla\cdot\vec{q} = 0 \quad (23)$$

*Angular Momentum Equation*

$$\rho k^2\nabla\cdot\frac{\vec{q}\vec{q}}{\varphi} = 2\tau\nabla\times\vec{q} \\ +(\alpha+\beta-\gamma)\varphi\nabla\left(\frac{1}{\varphi}\nabla\cdot\vec{g}\right) \\ +(\beta+\gamma)\nabla^2\vec{g} - 4\tau\vec{g} \quad (24)$$

*Linear Momentum Equation*

*For Granular Media:*

$$\rho\nabla\cdot\frac{\vec{q}\vec{q}}{\varphi} = -\varphi\nabla p^* + \nabla\cdot\vec{T}^* + 2\tau\nabla\times\vec{g} \\ -\frac{36(1-\varphi)^{\frac{2}{3}}(\eta_0 e^{\delta p^*} + \tau)}{d^2\varphi[1-(1-\varphi)^{\frac{1}{3}}][1-(1-\varphi)^{\frac{2}{3}}]}\vec{q} \\ -\frac{\rho C_d(1-\varphi)^{\frac{2}{3}}}{d[1-(1-\varphi)^{\frac{2}{3}}]^2}\vec{q}|\vec{q}| \quad (25)$$

*For Consolidated Media:*

$$\rho \nabla \cdot \frac{\vec{q}\vec{q}}{\varphi} = -\varphi \nabla p^* + \nabla \cdot \vec{T}^* + 2\tau \nabla \times \vec{g} - \frac{36\epsilon(\epsilon-1)}{(\varphi d)^2}(\eta_0 e^{\delta p^*} + \tau)\vec{q} - \frac{\rho C_d \epsilon(\epsilon-1)}{d\varphi(3-\epsilon)}\vec{q}|\vec{q}| \quad (26)$$

## 4 Conclusion

Intrinsic volume averaging was utilized in this work to derive equations governing the flow of an incompressible polar fluid with variable, pressure-dependent viscosity through isotropic porous media of variable porosity. Effects of the porous microstructure was quantified for granular and consolidated porous media using properties and characteristics of representative unit cell (RUC). Darcian drag and Forchheimer effects were identified in this work, and the dependence of Darcy's and Forchheimer's drag coefficients on pressure was discussed.

Future directions of this work include validation of the derived models; solving the model equations in an applied setting; and incorporating thermal effects in the models.

## Appendix 1:
## Volume Averaging Rules

A Representative Elementary Volume, REV, is a control volume that contains fluid and porous matrix in the same proportion as the whole porous medium. It is a control volume whose porosity is the same as that of the whole porous medium. Porosity, $\varphi$, is defined as the fraction of the pore volume, $V_\varphi$, to the bulk volume, $V$, of the medium, namely:

$$\varphi = V_\varphi / V \quad (A)$$

In terms of microscopic and macroscopic length scales, $l$ and $L$ respectively, the REV is chosen such that:

$$\ell^3 \ll V \ll L^3 \quad (B)$$

The volumetric phase average of a quantity $F$ is defined as:

$$<F> = \frac{1}{V}\int_{V_\varphi} F \, dV \quad (C)$$

and the intrinsic phase average (that is, the volumetric average of $F$ over the effective pore space, $V_\varphi$) is defined as:

$$<F>_\varphi = \frac{1}{V_\varphi}\int_{V_\varphi} F \, dV \quad (D)$$

The relationship between the volumetric phase average and the intrinsic phase average is obtained from equations (A), (C) and (D), and takes the form:

$$<F> = \varphi <F>_\varphi \quad (E)$$

Averaging theorems are then written in the following forms. Let $F$ and $H$ be volumetrically additive scalar quantities, $\vec{F}$ a vector quantity, and $c$ a constant (whose average is itself), then:

$$<cF> = c<F> = c\varphi<F>_\varphi \quad (i)$$

$$<\nabla F> = \varphi \nabla <F>_\varphi + \frac{1}{V}\int_S F°\vec{n}dS \quad (ii)$$

where $S$ is the surface area of the solid matrix in the REV that is in contact with the fluid, and $\vec{n}$ is the unit normal vector pointing into the solid. The quantity $F° = F - <F>$ is the deviation of the averaged quantity from its true (microscopic) value.

$$<F \mp H> = <F> \mp <H>$$
$$= \varphi<F \mp H>_\varphi = \varphi[<F>_\varphi \mp <H>_\varphi] \quad (iii)$$

$$<FH> = \varphi<FH>_\varphi \quad (iv)$$

$$= \varphi <F>_\varphi <H>_\varphi + \varphi <F°H°>_\varphi$$

$$<\nabla \cdot \vec{F}> = \nabla \cdot \varphi <\vec{F}>_\varphi + \frac{1}{V}\int_S \vec{F} \cdot \vec{n}\,dS \quad (v)$$

$$<\nabla \times \vec{F}> = \nabla \times \varphi <\vec{F}>_\varphi - \frac{1}{V}\int_S \vec{F} \times \vec{n}\,dS \quad (vi)$$

(*vii*)... Due to the no-slip condition, a surface integral is zero if it contains the fluid velocity vector explicitly.

(*viii*)... Due to the no-spin condition, a surface integral is zero if it contains the spin vector explicitly.

**Contribution of Individual Authors:** The authors equally contributed in the present research, at all stages from the formulation of the problem to the final findings and solution.

**Sources of Funding:** No funding was received for conducting this study.

**Conflict of Interest:**
The authors have no conflicts of interest to declare that are relevant to the content of this article.